\documentclass{kluwer}    % Specifies the document style.

\newdisplay{guess}{Conjecture}
\usepackage{epsfig}
\begin{document}                                                                                   
\begin{article}
\begin{opening}         
\title{Application of the global modal approach to the spiral galaxies}
\author{Natalia \surname{Orlova$^1$}}  
\author{Vladimir \surname{Korchagin$^1$}}
\author{Nobuhiro \surname{Kikuchi$^2$}}
\author{Shoken.M.\surname{Miyama$^3$}}
\author{Aleksei \surname{Moiseev$^4$}}
\runningauthor{Natalia Orlova}
\runningtitle{Application of the global modal approach to the spiral galaxies}
\institute{$^1$Institute of Physics, Rostov University, Stachki 194, Rostov-on-Don,
Russia, 344090;\\
 $^2$Earth Observation Research Center, National Space Development Agency of Japan,
1-8-10 Harumi, Chuo-ku, Tokyo 104-6023, Japan\\
$^3$National Astronomical Observatory, Mitaka, Tokyo 181-8588, Japan\\
$^4$Special Astrophysical Observatory, Nizhnij Arkhyz, Karachaevo-Cherkesia, 369167, Russia}
%\date{September 30, 2002}

\begin{abstract}
We have tested the applicability of the global modal approach 
in the density wave theory of spiral structure for a sample 
of spiral galaxies with measured axisymmetric background properties.
We report here the results of the simulations for four galaxies: NGC 488, NGC 628, NGC 1566, and NGC 3938.
Using the observed radial distributions for the stellar velocity 
dispersions and the rotation velocities we have 
constructed the equilibrium models for the galactic disks in each galaxy and 
implemented two kinds of stability analyses - the linear global analysis
and 2D-nonlinear simulations.  In general, the global modal approach is able to reproduce  the 
observed properties of the spiral arms in the galactic disks.
The growth of spirals in the galactic disks can be physically 
understood in terms of amplification by over-reflection 
at the corotation resonance. 
Our  results support the global modal approach 
as a theoretical explanation of  spiral structure in galaxies. 
\end{abstract}

\end{opening}           

%\section{Introduction}  
                    % Produces section heading.  Lower-level
                    % sections are begun with similar 
                    % \subsection and \subsubsection commands.

During the last 60 years many efforts have been made to reveal
the nature of the spiral structure in galaxies, but despite the progress made
in understanding the amplification mechanisms of the spiral arms and the role
of the spiral arms in the dynamics of galaxies, the answer is
still unclear. We do not even know if the spirals are a long-lived phenomenon,
or are short-lived and are regenerated many times during the galactic evolution.
According to the approach suggested by C.C. Lin and his collaborators,
spiral structure is a manifestation of quasi-stationary global modes in the galactic disks,
and are long-lived. In this picture,
the global modes are excited by over-reflection at corotation, and stay 'quasi-stationary' 
after being saturated at some level due to nonlinear effects of some sort.  
Another concept favors short-lived or  
recurrent spiral patterns which are developed in the galactic disks via swing-amplification
in a rotating disk with shear \cite{GoldreichandLynden}, or by an external
force \cite{Toomre}. 

A direct comparison of the theoretical models with the observed properties of
particular galaxies can help to discriminate between theories.
Typically, the procedure in the comparison studies
has relied on empirical estimates of the positions of the
corotation and Lindblad resonances, determined with the help of optical tracers in the
images of spiral galaxies.
The local dispersion relations were then used to calculate the shapes of spiral
patterns under additional assumptions about the radial behavior of Toomre's
stability parameter $Q$.
Obviously, assumptions made in such comparisons  reduce the predictive power of the theory.

We use a new approach to compare the observations with the spiral 
patterns theoretically predicted for the particular spiral galaxies.
By using measurements of the radial profiles of the stellar velocity dispersions in the galactic disks 
together with their rotation curves, we can reconstruct the disk's basic axisymmetric properties
without any additional assumptions. These data uniquely determine the
axisymmetric background equilibria of the disks in the particular galaxies, and can be used
for linear and nonlinear analyses aimed at modeling the spiral structure.

We use an approach similar to one we previously used 
to model the spiral structure in grand design spiral galaxy NGC 1566
\cite{Korchagin}.
Briefly, we construct the equilibrium models based on the
observed rotation curves and the velocity dispersion profiles in the galactic disks.
The radial dependence of the vertical velocity dispersion allows us to determine
the surface mass density distribution in the disks.
We then adopt a constant ratio of the vertical to the radial velocity
dispersion in the disk, varying it in the range \(0.7 \div 1\).
The stability properties of the particular galaxies have been studied
using a family of models within observational error bars.
We use a linear global modal analysis together with two-dimensional
nonlinear simulations of the dynamics of the galactic disks
seeded with noise perturbations. The linear global modal analysis
allows us to obtain the pattern speeds and growth rates of the
unstable modes together with the overall shape of the spirals.
The nonlinear simulations reproduce the dynamics of the unstable disks
seeded with the random perturbations, and allow us to follow the evolution
of the growing spirals up to their nonlinear saturation phase.
For all the models the results of linear and nonlinear analysis
are in an excellent agreement. 

\begin{figure}
\centerline{\includegraphics[width=9cm, angle=0]{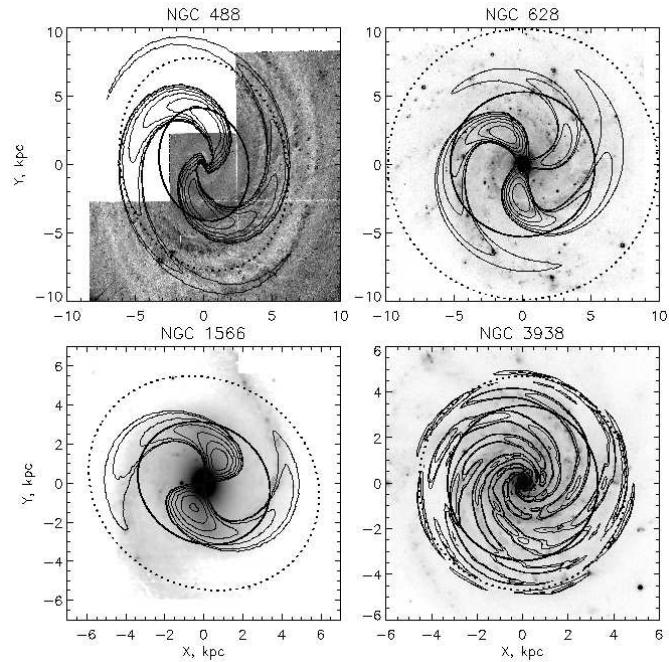}} 
\caption{The theoretical spiral patterns
(contour plots) superimposed onto the images of four spiral galaxies
NGC 488, NGC 628, NGC 1566 and NGC 3938. The solid line shows the positions
of corotation resonance for the most unstable mode.
The dotted lines indicate the positions of outer Lindblad resonance.} 
\end{figure}

A comparison of the theoretically predicted  spiral patterns 
with the optical/NIR images of spiral galaxies is shown in Figure 1.
In general, the global modal approach reproduces qualitatively
well the observed morphologies of spiral galaxies.
The observed spiral patterns can be explained as
a superposition of the unstable global modes growing in the disks
from the arbitrary noise perturbations.
The galaxies with more massive disks usually develop
two-armed patterns, while the galaxies which have
low mass disks and smaller stellar velocity dispersion
in the central regions tend to develop a patchy multi-armed structure.

{\bf NGC 488.} Upper left panel of Figure 1 shows the theoretical
spiral pattern superimposed onto the HST B-band image of this galaxy
after subtracting the bulge contribution to the surface brightness distribution.
The early-type Sb galaxy NGC 488 has a regular tightly wound two-armed
spiral structure which can be traced close to the central region.
The theoretical modal analysis and nonlinear simulations show that the dynamics of the disk
in NGC 488 is governed by the most unstable two-armed spiral.
The theoretical spiral pattern is, however, more open compared to the
observed spiral, especially in the central regions of the disk.

{\bf NGC 1566.} The Sc galaxy NGC 1566 has an open grand design spiral pattern.
The lower left panel of Figure 1 shows the theoretical spiral pattern
superimposed onto the K-band image of NGC 1566 (Mulchaey et al. 1997). 
The pattern is dominated by a two-armed spiral mode during the
overall evolution.
There is a good correspondence between the theoretical prediction
and observations.

{\bf NGC 628.} This Sc galaxy has an asymmetric two-armed pattern with 
the northern arm split into two arms. A linear modal analysis
shows the two unstable modes $m$ = 2, and $m$ = 3 of comparable growth rates.
The upper right panel of Figure 1 shows a theoretical spiral pattern
taken from the nonlinear simulations superimposed onto the B-band
image taken by Larsen \& Richtler (1999). The theory
qualitatively correctly predicts the morphology of the spiral
pattern showing that the observed pattern is indeed a
superposition of two unstable modes. However, the winding of the observed spiral pattern
is higher in the central regions compared to the theoretical prediction
which might be caused by the uncertainties in the measurements of the
velocity dispersion and the rotation curve in the central regions of the galaxy.

{\bf NGC 3938.} This Sc galaxy has a multi-armed structure. A linear modal
analysis shows that the disk of this galaxy is most unstable toward  $m$ = 3 and $m$ = 4
modes. This results in a multi-armed spiral pattern,
which is confirmed by our nonlinear simulations.

\centerline {\bf {Conclusions}}

The global modal approach correctly predicts
the shapes of the spiral patterns in galaxies of different morphological types.
The growth rates of the governing modes can be physically understood in terms of
amplification at the corotation resonance (Drury 1980).
Our results thus strongly support the global modal approach 
as a theoretical explanation of  spiral structure in galaxies, and can be
considered as indirect evidence the long-lived spiral patterns
in galactic disks.

\end{article}
\end{document}